\documentclass[5p,times,twocolumn]{elsarticle}
\usepackage{graphicx}
\usepackage{txfonts}
\usepackage{epsfig}
\journal{Nuclear Instruments and Methods in Physics Research A}

\begin{document}

\begin{frontmatter}

\title{First results about on-ground calibration of the Silicon Tracker for the AGILE satellite}

%\author{P.W. Cattaneo\corref{cor1}} \cortext[cor1]{Speaker} 
%        \ead{Paolo.Cattaneo@pv.infn.it }

%\address {\emph{INFN Pavia, Via A. Bassi 6, I-27100 Pavia, Italy }} 

\author[label7]{P.W. Cattaneo \corref{cor1}} \cortext[cor1]{Corresponding author}
        \ead{Paolo.Cattaneo@pv.infn.it }
\author[label1]{A. Argan}
\author[label7]{F. Boffelli}
\author[label5]{A. Bulgarelli}
\author[label24]{B. Buonomo}
\author[label3,label4]{A.W. Chen}
\author[label1,label2]{F. D'Ammando}
\author[label2,label4]{T. Froysland}
\author[label5]{F. Fuschino}
\author[label8]{M. Galli}
\author[label5]{F. Gianotti}
\author[label3]{A. Giuliani} 
\author[label6]{F. Longo}
\author[label5]{M. Marisaldi }
\author[label24]{G. Mazzitelli}
\author[label20]{A. Pellizzoni} 
\author[label12]{M. Prest}
\author[label1]{G. Pucella}
\author[label24]{L. Quintieri}
\author[label7]{A. Rappoldi}
\author[label1,label2]{M. Tavani} 
\author[label5]{M. Trifoglio }
\author[label1]{A. Trois }
\author[label24]{P. Valente}
\author[label6]{E. Vallazza }
\author[label22]{S. Vercellone} 
\author[label3]{A. Zambra }
\author[label6]{G. Barbiellini}
\author[label3]{P. Caraveo} 
\author[label1]{V. Cocco}
\author[label1]{E. Costa}
\author[label1]{G. De Paris}\author[label1]{E. Del Monte}
\author[label5]{G. Di Cocco}\author[label1]{I. Donnarumma}
\author[label1]{Y. Evangelista} \author[label1]{M. Feroci}
\author[label4,label18]{A. Ferrari} \author[label3]{M. Fiorini}
\author[label5]{C. Labanti}\author[label1]{I. Lapshov}
\author[label1]{F. Lazzarotto}
\author[label9]{P. Lipari}
\author[label10]{M. Mastropietro}
\author[label3]{S. Mereghetti}
\author[label5]{E. Morelli}\author[label6]{E. Moretti}
\author[label11]{A. Morselli}\author[label1]{L. Pacciani}
\author[label3]{F. Perotti} 
\author[label1,label2,label11]{G. Piano}
\author[label2,label11] {P. Picozza}
\author[label12]{M. Pilia}\author[label1]{G. Porrovecchio}
\author[label13]{M. Rapisarda }
\author[label1]{A. Rubini} \author[label1,label2]{S. Sabatini }
\author[label1]{P. Soffitta}
\author[label2,label11]{E. Striani }
\author[label1,label2]{V. Vittorini }
\author[label9]{D. Zanello }
\author[label14]{S. Colafrancesco } \author[label14]{P. Giommi}
\author[label14]{C. Pittori} \author[label14]{P. Santolamazza }
\author[label14]{F. Verrecchia} \author[label]{}
\author[label15]{L. Salotti }

\address[label1] {INAF/IASF-Roma, I-00133 Roma, Italy}\address[label2]  
{Dip. di Fisica, Univ. Tor Vergata, I-00133 Roma,Italy}  
\address[label3] {INAF/IASF-Milano, I-20133 Milano, Italy}
\address[label4] {CIFS-Torino, I-10133 Torino, Italy}
\address[label5] {INAF/IASF-Bologna, I-40129 Bologna, Italy}
\address[label6] {INFN Trieste, I-34127 Trieste, Italy} 
\address[label7] {INFN-Pavia, I-27100 Pavia, Italy}
\address[label8] {ENEA-Bologna, I-40129 Bologna, Italy}
\address[label9] {INFN-Roma La Sapienza, I-00185 Roma, Italy}
\address[label10] {CNR-IMIP, Roma, Italy}
\address[label11] {INFN Roma Tor Vergata, I-00133 Roma, Italy} 
\address[label12] {Dip. di Fisica, Univ. Dell'Insubria, I-22100 Como, Italy}
\address[label13] {ENEA Frascati,  I-00044 Frascati (Roma), Italy}
\address[label14] {ASI Science Data Center, I-00044 Frascati(Roma), Italy} 
\address[label15] {Agenzia Spaziale Italiana, I-00198 Roma, Italy} 
\address[label16] {Osservatorio Astronomico di Trieste, Trieste, Italy} 
\address[label18] {Dip. Fisica, Universit\'a di Torino, Turin, Italy}
\address[label20] {INAF-Osservatorio Astronomico di Cagliari, 
localita' Poggio dei Pini, strada 54, I-09012 Capoterra, Italy} 
\address[label22] {INAF-IASF Palermo, Via Ugo La Malfa 153, I-90146  Palermo, Italy}
\address[label23] {Dip. Fisica Univ. di Trieste, I-34127 Trieste, Italy} 
\address[label24] {INFN Lab. Naz. di Frascati, I-00044 Frascati(Roma), Italy} 

\begin{abstract}
    
The AGILE scientific instrument has been calibrated with a tagged $\gamma$-ray beam
at the Beam Test Facility (BTF) of the INFN Laboratori Nazionali di Frascati (LNF).
The goal of the calibration was the measure of the Point Spread Function (PSF) as a function 
of the photon energy and incident angle and the validation of the Monte Carlo (MC) simulation
of the silicon tracker operation.
The calibration setup is described and some preliminary results are presented.
\end{abstract}


\end{frontmatter}

\section{The AGILE mission}

\begin{figure}[!ht]
\begin{center}
\includegraphics[width=5cm]{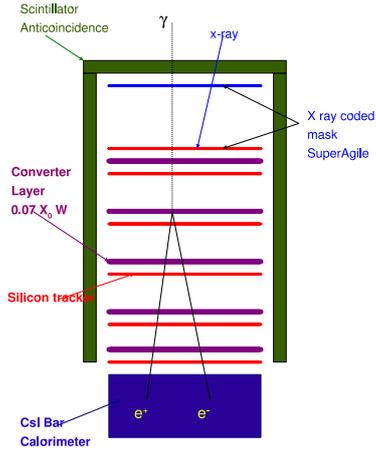}
\end{center}
\caption{A schematic view of the AGILE scientific instrument.}
\label{agidraw}
\end{figure}

AGILE (Astro-rivelatore Gamma a Immagini LEggero) is a Small Scientific
Mission of the Italian Space Agency (ASI) launched on April 2007 and
dedicated to high-energy astrophysics \cite{agimis}.
The AGILE satellite is designed to detect and image photons in the 
18 - 60 keV, 30 MeV - 50 GeV and 350 keV - 100 MeV energy bands with excellent 
spatial resolution, timing capability, and large field of view.\\
AGILE is the most compact ($\approx 0.25 m^3$), light (120 kg for the instrument,
350 kg for the whole satellite) and low power ($\approx 60 W$) scientific instrument
ever developed for high-energy astrophysics.\\
The AGILE scientific payload (shown in Fig.\ref{agidraw}) consists of three detectors 
with independent detection capability.
The Gamma-Ray Imaging Detector (GRID) consists of a Si-W converter-tracker \cite{st1}
sensitive in the $\gamma$-ray energy range 30 MeV - 50 GeV,
a shallow ($1.5\, X_0$ on-axis) CsI Calorimeter \cite{minical} and a segmented 
AntiCoincidence system based on plastic scintillators \cite{ac}.\\
In addition to the GRID, a coded-mask hard X-ray imaging system (SuperAGILE), 
made of a Si detector plane and a W mask, ensures coverage in the range 
$18-60 \mathrm{keV}$ \cite{superagile}.\\ 
The AGILE main feature is the combination of two co-aligned imaging detectors 
(SuperAGILE and GRID) sensitive in the hard X-ray and in the $\gamma$-ray ranges 
with large field of view ($\approx 1.0\mathrm{sr}$ and $\approx 2.5 \mathrm{sr}$ 
respectively).\\
Moreover the CsI MiniCalorimeter (MCAL) can operate in stand alone "burst mode" 
in the $350\,\mathrm{keV}- 100\,\mathrm{MeV}$ range to detect GRB.\\
On ground and subsequently on flight calibrations of a detector are essential 
to the interpretation of its results. The purpose of the calibration of a 
scientific instrument is to reproduce, under 
controlled condition, the detector response in operation.\\
This paper describes the on-ground calibration of the silicon tracker and some results
on the instrument performances derived by it.

%%%%%%%%%%%%%%%%%%%%%%%%%%%%%%%%%%%%%%%%%%%%%%%%%%%%%%%%%%%%%%%%%%%%%%%%%%%%%%%%%%%%%%%
%%%%%%%%%%%%%%%%%%%%%%%%%%%%%%%%%%%%%%%%%%%%%%%%%%%%%%%%%%%%%%%%%%%%%%%%%%%%%%%%%%%%%%%%%

\section{The Silicon Tracker}

The core of the GRID is the Silicon Tracker (ST) that converts the $\gamma$-rays and 
measures the trajectories of the resulting $e^+/e^-$ pairs \cite{st1}-\cite{sttb} 
%(see Fig.\ref{agidraw}).
The ST consists of 12 trays with distance between middle-planes equal to $1.9\,\mathrm{cm}$ 
optimized by simulation. The first 10 trays consist of a W converter layer 
$245\,\mu \mathrm{m}$ thick followed by pairs of single sided Si microstrip planes 
with strips orthogonal to each other to provide three dimensional points
(corresponding to a total thickness $0.01(Si)+0.07(W) X_0$ ). The last two trays have 
no W converter layers since the GRID trigger logic requires at least three contiguous Si planes.\\ 
The detector unit is a $9.5\times 9.5 \mathrm{cm}^2$ tile, $410\,\mu m$ thick
with strip pitch $121\,mu m$. Four tiles bonded together form a 'ladder'.
Every ST plane consists of four ladders.\\
Only every second strip is readout to limit the power consumption. The non readout strips 
contribute to the resolution through the principle of capacitive charge division.\\
Each ladder is read-out by three TAA1 ASICs, each operating 128 channels at low noise,
low power configuration ($< 400 \mu$W/channel), self-triggering ability and analog readout.\\
The ST position resolution is below $40 \mu \mathrm{m}$ for a large range of particle incidence
angles \cite{sttb}.

\subsection{The GRID simulation}

The GRID as mounted on the spacecraft and as installed in the test beam is simulated using 
the GEANT 3.21 package \cite{geant}. This package provides for a detailed simulation of the materials
and describes with high precision the passage of particles through matter including the production of 
secondary particles.\\
%In a microstrip detector with floating strips a critical aspect of the simulation is the sharing 
%of the charge collected on the strips to the readout channels. These sharing coefficients are 
%estimated by test beam data as detailed in \cite{sttb} to reproduce the data.
The simulation output is formatted to be readable by the reconstruction programs used 
for the analysis of in-flight data. 

\subsection{Direction and Energy Reconstruction}

The $\gamma$-ray direction reconstruction is obtained from the identification and the
analysis of the $e^+/e^-$ tracks stemming from the conversion vertex. Each microstrip 
silicon plane measures separately the X and Y hit coordinates.\\
The first step of the event analysis requires to find two tracks among the
possible associations of the hits detected by the ST layers.\\
The second step consists in fitting the track trajectories through the hits
accounting for the presence of energy loss and multiple scattering.
These steps are performed separately for the X and Y coordinates producing
four tracks, two for each projection. The three dimensional direction is obtained 
requiring a correct association of the two projections of each track.\\
The track parameters are fitted by a Kalman filter smooth algorithm \cite{frukal}. 
A special implementation of the filter \cite{giukal} exploits the measurement 
of the angular scattering of the $e^\pm$ due to the interactions with the material 
to estimate the track energies. 
Combining the track energies the $\gamma$-ray energy is estimated.

\section{The $\gamma$-ray Calibrations}

\subsection{Calibration goals}

The goal of the calibration is to estimate the instrument response
function by exposing it to a $\gamma$-ray beam with energy and direction
known to an accuracy better than the resolving power of the instrument.

The required accuracy of ST is driven by its use during the AGILE mission: 
the systematic errors introduced by the calibration should be smaller than 
the statistic errors expected from a bright celestial source. 

The detector properties to be evaluated by the calibration are: 
the detection efficiency, the angular resolution, the energy resolution.  
In this paper we concentrate on evaluating the Point Spread Function (PSF)
as a function of the $\gamma$-ray energy and incident angle.

The calibration is also intended to validate the MC simulation program.
This simulation will be required to complement the calibration data in 
the untested parts of parameter space. In particular the information above 
the maximum energy available at BTF can be obtained only through the simulation.

The calibration is designed to cover a wide range of the geometries 
and conditions realized in space. The ST was calibrated at the INFN LNF 
in the period 2-20 November 2005, thanks to the collaboration 
between the AGILE Team and INFN-LNF. 

\subsection{Calibration strategy}

To meet the calibration accuracy requirements, we have determined the
number of $\gamma$-rays required for the calibration of AGILE, taking into account
the photon fluence of a characteristic $\gamma$-rays reference source as Vela.

With a cover-up efficiency of 50$\%$ and an effective area of $\approx 500
\mathrm{cm}^2$, the number of counts estimated is about 
10$^4$ for E $>$ 100 MeV, after two months of observation. 
The requirement on the number of calibration photons detected by the GRID 
is $4 \times 10^5$ for $E_\gamma > 30\,\mathrm{MeV}$, $4\times 10^4$ for
$E_\gamma > 100\,\mathrm{MeV}$.      

\subsection{Calibration set up}

\subsubsection{The Beam Test Facility}

For the ST calibration we used the Beam Test Facility (BTF) in the Frascati 
DA$\Phi$NE collider complex, which includes a LINAC at high $e^-/e^+$ currents,
an accumulator of $e^-/e^+$ and two accumulation rings at 510 MeV.

%%%%%%%%%%%%%%%%%%%%%%%%%%%%%%%%%%%%%%%%%%%%%%%%%%%%%%%%%%%%%%%%%%%%%%%%%%%%%%%%%%%%%%%
%\begin{figure}[!ht]
%\begin{center}
%\includegraphics[height=6cm,angle=270]{fascioFrascati.eps}
%\end{center}
%\caption{The transfer line Beam Test Facility (BTF).}
%\label{linea-fascio}
%\end{figure}
%%%%%%%%%%%%%%%%%%%%%%%%%%%%%%%%%%%%%%%%%%%%%%%%%%%%%%%%%%%%%%%%%%%%%%%%%%%%%%%%%%%%%%%%%

The $e^+/e^-$ beam from the LINAC is led into the accumulation ring to be subsequently 
injected in the principal ring. When the beam is not transferred in the accumulator, it 
can be transported from the LINAC in the test beam area 
through a dedicated transfer line: the BTF line. 
%(Fig.~\ref{linea-fascio})
The BTF provides a collimated beam of $e^-/e^+$ in the energy range
20-800 MeV with a pulse rate of 50 Hz. The pulse duration can vary from 1 to
10 ns and the number of particles for bunch can range from 1 to $10^5$.\\
We operated with energy beam of 463 MeV and a pulse duration of 2 ns.
%The BTF can be operated in two ways
%\begin{itemize}
%\item a LINAC mode operating only when DA$\Phi$NE is off with a maximum energy of 750 MeV and 
%an efficiency $\approx 0.9$
%\item a DA$\Phi$NE mode operating only DA$\Phi$NE is on with a fixed energy of 750 MeV and 
%an efficiency $\approx 0.2$
%\end{itemize}

\subsubsection{Target }

$\gamma$-rays were produced by Bremsstrahlung in a thin Silicon target; subsequently a 
magnet bent away the $e^-$ while the $\gamma$-rays could impinge on the GRID.\\
The target is constituted by two pairs of silicon microstrip single sided detectors of
$8.75 \times 8.75 \mathrm{cm}^2$ and $0.41 \mathrm{mm}$ thick, including 384 strips with 
$228 \mu m$ pitch. The target measures the passage of the $e^-$ 
and cause the emission of Bremsstrahlung $\gamma$-ray.

%%%%%%%%%%%%%%%%%%%%%%%%%%%%%%%%%%%%%%%%%%%%%%%%%%%%%%%%%%%%%%%%%%%%%%%%%%%%%%%%%%%%%%%
\begin{figure}[!ht]
\begin{center}
\includegraphics[height=6cm,angle=0]{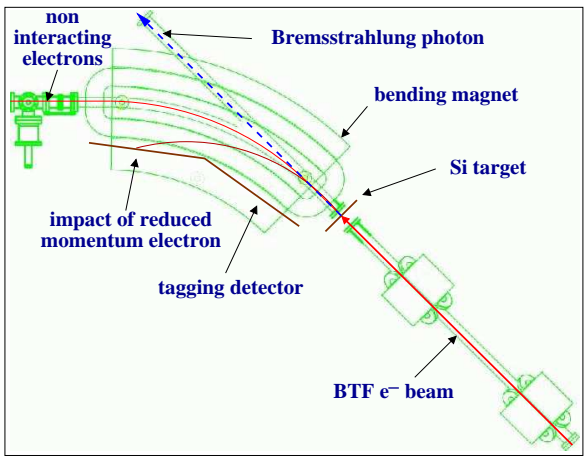}
\end{center}
\caption{A schematic of the $\gamma$-ray line: the target, the bending magnet and the PTS.}
\label{tagger}
\end{figure}
%%%%%%%%%%%%%%%%%%%%%%%%%%%%%%%%%%%%%%%%%%%%%%%%%%%%%%%%%%%%%%%%%%%%%%%%%%%%%%%%%%%%%%%%%

\subsubsection{Tagging system}

Our Team developed and installed in the BTF area a Photon Tagging System (PTS)
for the detection of the particles interacting with the target. 
The $e^-$ are tagged using microstrip Si detectors located on the internal walls
of the bending dipole magnet (see Fig.\ref{tagger}). Depending on the energy loss 
in the target, the $e^-$ impinge on different strips.
The correlation of the measurements of the $e^-$ by the target Si planes 
and by the PTS tags the photon; the position on the PTS measures the photon energy.\\
The PTS operates in self-trigger mode, i.e. it is readout independently from
the GRID. This point has a great relevance for the following analysis.\\
During the 18 days of calibration, about $2,10^5$ tagged $\gamma$-rays were produced,
of which $\approx 40\%$ interacted with the GRID. 

\subsection{Instrument Ground Support Equipment (GSE)}

We developed and installed specific equipment  required to coordinate and, whenever 
possible, automate the instrument management and the data gathering and analysis as 
required by the calibration procedures. The Mechanical Ground Support Equipment 
(MGSE) \cite{gianotti} hosts the payload and allows the precise motorized 
translations and manual rotations of the detector volume in front of the beam. 
In near real time, the Science Console (SC) \cite{bulgarelli} archives all the 
instrument data and performs the quick look to check the instrument behaviour. 
It is also in charge of  producing the energy histogram of the PTS data to verify the
actual statistic of the PTS measurement and decide the measurement duration.

\subsection{Trade-off on the number of e$^-$/bunch}

The GRID performance should be evaluated in a 'single-photon' regime without 
simultaneous multi-photon interactions. 
Multiple photon events are not representative of astrophysical conditions 
and may introduce a significant bias in the measurement.

The best configuration was with 1 e$^-$/bunch, but considering the time 
available for calibration and to obtain a higher efficiency
we adopted 3 e$^-$/bunch.

\subsection{Simulation}

The overall system including the beam terminal section, the target, the bending 
magnet, the PTS and the GRID are simulated in detail using GEANT 3.21 package 
\cite{geant}.\\
That allows a direct comparison between the resolutions measured in simulated and 
real data providing a check of the quality of the MC simulations.\\
A significant improvements of the comparison between data and MC were obtained 
by overlapping a uniform flux of low energy $\gamma$-rays to the Bremsstrahlung 
$\gamma$-ray. These $\gamma$-ray represents a background that cannot be precisely 
and is tuned to match the experimental data.

\section{Data Analysis}
     
\subsection{Data Samples}

The $\gamma$-ray beam was directed to the ST at fixed $\theta$ and $\phi$
with respect to the detector planes. The beam spot on the detector is small, 
$\approx 2-3\,\mathrm{mm}$ as measured in the target. 
Yet, ideally the photon beam should illuminate uniformly the ST.\\
An approximated uniform illumination is obtained by translating the 
GRID with respect to the beam on a run by run basis. 
The beam impinge in four to eight different positions per orientation, called spills.\\
Data were collected for different combinations of $\theta = 0^\circ, 30^\circ, 50^\circ$ 
and $\phi = 0^\circ, 45^\circ, 135^\circ, 225^\circ, 270^\circ, 315^\circ$.\\
Runs for different spills and same orientation are grouped together. 
Also runs for different $\phi$ and same $\theta$ are 
grouped together after having verified that they are compatible. 
 
\subsection{The GRID trigger}

The GRID trigger for AGILE operation is described in \cite{agimis}.
The relevant point for this calibration is the following: the GRID is self-triggering,
that is no external signal nor phase locking with the accelerator is present.\\
During the calibration the in flight trigger configuration was 
active except for the AntiCoincidence veto that was turned off. This choice was imposed
by the high rate of background induced hits in the experimental hall that was reducing the 
live time to an unacceptable level.
In this configuration the triggered events were contaminated by charged particles crossing the
AntiCoincidence panels.

%%%%%%%%%%%%%%%%%%%%%%%%%%%%%%%%%%%%%%%%%%%%%%%%%%%%%%%%%%%%%%%%%%%%%%%%%%%%%%%%%%%%%%%%%%%%%%%%
\begin{figure}[!ht]
%\vspace{7cm}
\begin{center}
\includegraphics[width=5cm,height=5cm]{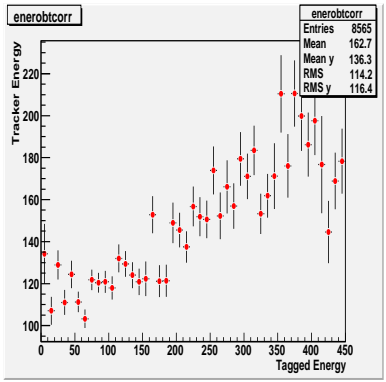}
\end{center}
\caption{Relation between ST and PTS energy}
\label{enerobtcorr}
\end{figure}
%%%%%%%%%%%%%%%%%%%%%%%%%%%%%%%%%%%%%%%%%%%%%%%%%%%%%%%%%%%%%%%%%%%%%%%%%%%%%%%%%%%%%%%%%%%%%%%%%%%

%%%%%%%%%%%%%%%%%%%%%%%%%%%%%%%%%%%%%%%%%%%%%%%%%%%%%%%%%%%%%%%%%%%%%%%%%%%%%%%%%%%%%%%%%%%%%%%%
\begin{figure}[!ht]
\begin{center}
\includegraphics[height=5cm,width=5cm]{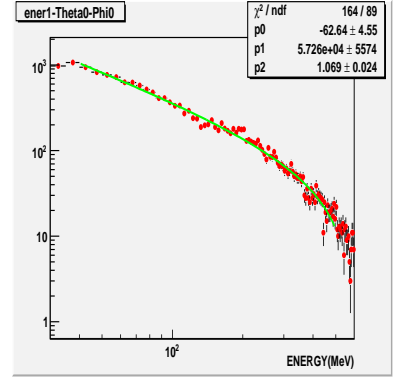}
\end{center}
\caption{$E_\gamma$ measured from the GRID fitted with $p_0+p_1E^{-p_2}$ for $\theta=0^\circ$ 
and $\phi=0^\circ$}
\label{EGRID}
\end{figure}
 
\subsection{Event reconstruction: the filter}

The event reconstruction program, called filter, selects events with 
a $\gamma$-ray converting in a $e^+/e^-$ pair.
The events are expected to have two tracks from a vertex within the ST. 
The kinematic of the event is reconstructed applying  a Kalman filter.
The energy of the tracks are estimated by the multiple scattering in the ST planes.\\
In addition the filter returns a flag assigning an estimation for the event being 
a $\gamma$-ray or background.\\
There are four flags tagged as G (gamma), L (limbo), P (particle), S (single).
The events flagged as G satisfies very strict requirements to be a converted $\gamma$-ray.\\
Those flagged as limbo are possible but not certain $\gamma$-ray.\\
Those flagged as particle are estimated to be particle crossing the ST (e.g. cosmic muons).\\
Those flagged as single are estimated to be single particles from a vertex within the ST.\\
The Point Spread Function (PSF) of the ST ideally should be studied with a sample of 
G events having a minimal background contamination. However, when the PSF is studied for 
each orientation and versus the energy, sufficient statistic is required.\\
The unselected triggered events at $\theta=0^\circ$ are more than $3\,10^6$ events. 
The flag fractions are: G(2.3\%), L(48.3\%) P(38.6\%), S(10.8\%). 
The striking feature is the low fraction of G events. This is a feature of the high background 
environment present in the BTF. In this regard the in flight environment is much cleaner and the 
fraction of G events is much higher.
 
\subsection{Event selection for data: the PTS}

The PTS can be used for two different but related purposes: 1) as an off-line trigger to identify
the emission of Bremsstrahlung $\gamma$-ray in the target and 2) as a device to measure the $\gamma$-ray 
energy regardless from the ST.\\
The PTS and GRID events are paired off-line exploiting the event times measured in both
devices up to $1 \mu s$ precision.\\
The PTS is required to have a very clean signal to reduce multi-$\gamma$-ray events and various 
background sources. That implies a low efficiency as off-line trigger. In the 
configuration $\theta=0^\circ$ the tagged events are only 23596. The fractions of events in the four 
flags are: G(5.2\%), L(44.5\%) P(30.3\%), S(20.0\%). There is a significant increase in the
fraction of G events that nevertheless remain a small fraction. The same pattern is present for the
other orientations.\\
The other task of the PTS is the measurement of $E_\gamma$. This is obtained 
calibrating with the MC the relation between $E_\gamma$ and the position of interaction of the 
$e^-$ on the PTS. A close relation between $E_\gamma$ measured by the PTS and by the ST
is expected. Fig.\ref{enerobtcorr} shows the profile plot of ST energy
versus PTS energy. The correlation is significant, but the spread is large and the 
linearity is poor. The GRID energy resolution cannot be evaluated precisely with this method.\\
The quality of the GRID energy measurement can be estimated from Fig.\ref{EGRID} where the 
$E_\gamma$ spectrum measured by the GRID is fitted with the function $p_0+\frac{p_1}{E^{p_2}}$ 
as expected for a Bremsstrahlung spectrum. The limited distortion indicates that 
the GRID energy resolution and the energy dependence of the efficiency do not alter 
significantly the Bremsstrahlung spectrum.

%%%%%%%%%%%%%%%%%%%%%%%%%%%%%%%%%%%%%%%%%%%%%%%%%%%%%%%%%%%%%%%%%%%%%%%%%%%%%%%%%%%%%%%%%%%%%%%%%%%
%
%\subsection{PSF for tagged real data}
% 
%\subsection{Event selection for MC: the PTS}
% 
%\subsection{Comparison of real data and MC: the PTS}
 
\subsection{Event selection for data: the phase approach}

\begin{figure}[!ht]
\begin{center}
\includegraphics[height=5cm]{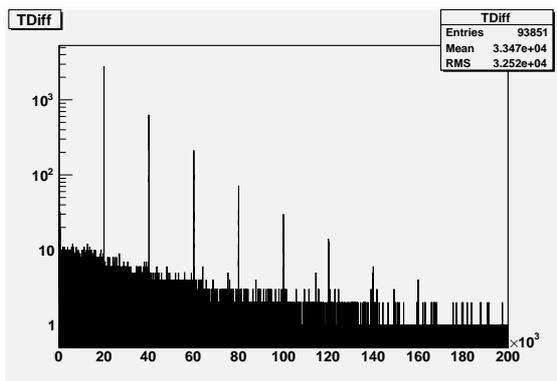}
\end{center}
\caption{Time difference between consecutive events in a run}
\label{TDiff}
\end{figure}

\begin{figure}[!ht]
\begin{center}
\includegraphics[height=6cm,width=6cm]{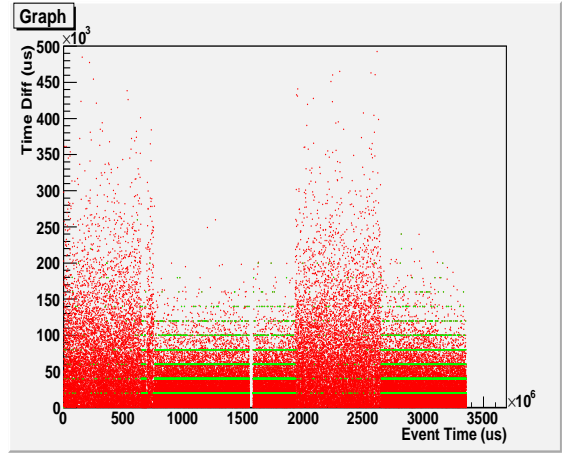}
\end{center}
\caption{Time difference between consecutive events versus time for a run. Events in phase
are marked in lighter color}
\label{TDiffvsTime}
\end{figure}

A drawback of the PTS approach is the low efficiency $\approx 1\%$.
If high quality reconstruction (flag G) is required, the number of events available 
for the PSF determination in any given configuration may become very small.\\
An alternative approach consists in exploiting the BTF bunched periodicity at
$50\,Hz$. The intra spill period is $T_{BTF}=20\,ms$.\\
That implies that subsequent beam related $\gamma$-ray are spaced in time of multiples 
of $T_{BTF}$, {\it in phase} with the 
beam period. The event time on the GRID is measured with a resolution of 
$1\,\mu s$ that defines the precision of the selection.\\ 
In Fig.\ref{TDiff} the distribution of time differences between consecutive events is 
displayed, showing high peaks in correspondence of $T_{BTF}$ multiples.\\
Fig.\ref{TDiffvsTime} presents the event phase versus the event time. Events are 
{\it in phase} if the time difference between consecutive events is an integer 
multiple of $T_{BTF}$ within $100\,\mu\mathrm{s}$. These events are marked in 
lighter color.\\
Another prominent feature of Fig.\ref{TDiffvsTime} can be interpreted as follows: there 
are time intervals (approximately 0-500 s and 1800-2700 s) showing no accumulation of events
{\it in phase}. That is a sign of beam off time when the GRID measures only beam unrelated 
background. Outside these intervals there is an accumulation of events {\it in phase} with
decreasing numbers when the number of $T_{BTF}$ increases.
Restricting to the beam on intervals, the fraction of events {\it in phase} is $\approx 25\%$.

%%%%%%%%%%%%%%%%%%%%%%%%%%%%%%%%%%%%%%%%%%%%%%%%%%%%%%%%%%%%%%%%%%%%%%%%%%%%%%%%%%%%%%%%%%%%%%%%
%%%%%%%%%%%%%%%%%%%%%%%%%%%%%%%%%%%%%%%%%%%%%%%%%%%%%%%%%%%%%%%%%%%%%%%%%%%%%%%%%%%%%%%%%%%%%%%%%%%

%Events are defined to be 'in phase' if the time elapsed from the previous event is a integer 
%multiple of $T_{BTF}$ within $100\,\mu s$. These events are marked as green and the others as red.\\
%The patterns in Fig.\ref{TDiff} can be interpreted as follows: there are significant time intervals 
%(approximately the intervals 0-500 ms and 1800-2700 ms) showing no significant accumulation of 
%events 'in phase', that is a sign of beam off when the GRID measures only beam unrelated background;
%in the other intervals there is an accumulation of events 'in phase' with decreasing amplitudes when
%the number of $T_{BTF}$ increases. Restricting to the beam on intervals, the fraction of events 
%'in phase' is $\approx 25\%$.

%%%%%%%%%%%%%%%%%%%%%%%%%%%%%%%%%%%%%%%%%%%%%%%%%%%%%%%%%%%%%%%%%%%%%%%%%%%%%%%%%%%%%%%%%%%%%%%%
%%%%%%%%%%%%%%%%%%%%%%%%%%%%%%%%%%%%%%%%%%%%%%%%%%%%%%%%%%%%%%%%%%%%%%%%%%%%%%%%%%%%%%%%%%%%%%%%%%%

Events {\it in phase} can be selected regardless the presence of tagging to enhance the available 
statistic and the PSF can be estimated by these samples.\\
If this approach is correct all tagged events are expected to be {\it in phase}.
That is the case confirming the validity of the approach.

%%%%%%%%%%%%%%%%%%%%%%%%%%%%%%%%%%%%%%%%%%%%%%%%%%%%%%%%%%%%%%%%%%%%%%%%%%%%%%%%%%%%%%%%%%%%%%%%
\begin{figure}[!ht]
\begin{center}
\includegraphics[height=6cm,width=6cm]{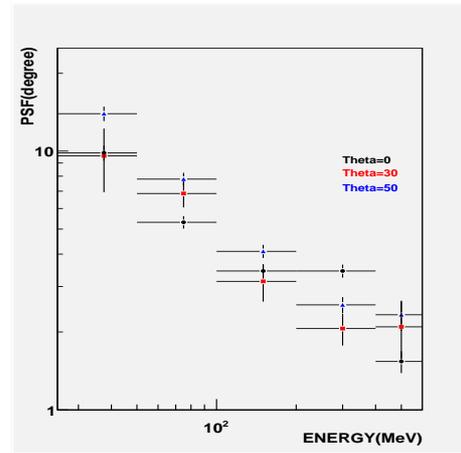}
\end{center}
\caption{PSF (68\%) versus energy for MC.}
\label{PSF68MC}
\end{figure}
%%%%%%%%%%%%%%%%%%%%%%%%%%%%%%%%%%%%%%%%%%%%%%%%%%%%%%%%%%%%%%%%%%%%%%%%%%%%%%%%%%%%%%%%%%%%%%%%%%%

%%%%%%%%%%%%%%%%%%%%%%%%%%%%%%%%%%%%%%%%%%%%%%%%%%%%%%%%%%%%%%%%%%%%%%%%%%%%%%%%%%%%%%%%%%%%%%%%
\begin{figure}[!ht]
\begin{center}
\includegraphics[height=6cm,width=6cm]{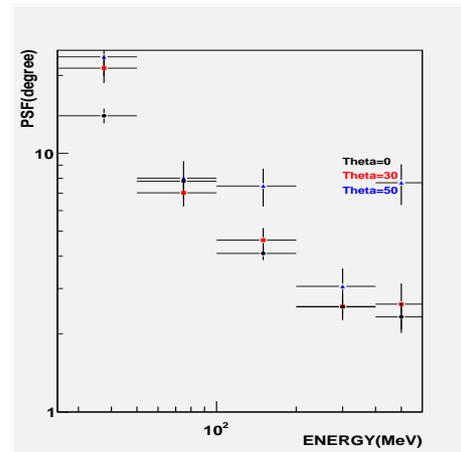}
\end{center}
\caption{PSF (68\%) versus energy for data.}
\label{PSF68Data}
\end{figure}
%%%%%%%%%%%%%%%%%%%%%%%%%%%%%%%%%%%%%%%%%%%%%%%%%%%%%%%%%%%%%%%%%%%%%%%%%%%%%%%%%%%%%%%%%%%%%%%%%%%

%%%%%%%%%%%%%%%%%%%%%%%%%%%%%%%%%%%%%%%%%%%%%%%%%%%%%%%%%%%%%%%%%%%%%%%%%%%%%%%%%%%%%%%%%%%%%%%%
\begin{figure}[!ht]
\begin{center}
\includegraphics[height=6cm,width=6cm]{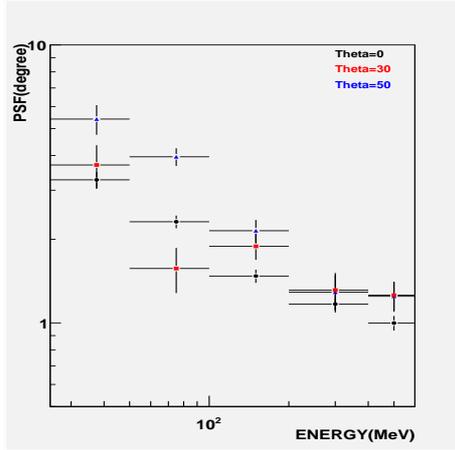}
\end{center}
\caption{Gaussian PSF versus energy for MC.}
\label{PSFMC}
\end{figure}
%%%%%%%%%%%%%%%%%%%%%%%%%%%%%%%%%%%%%%%%%%%%%%%%%%%%%%%%%%%%%%%%%%%%%%%%%%%%%%%%%%%%%%%%%%%%%%%%%%%

%%%%%%%%%%%%%%%%%%%%%%%%%%%%%%%%%%%%%%%%%%%%%%%%%%%%%%%%%%%%%%%%%%%%%%%%%%%%%%%%%%%%%%%%%%%%%%%%
\begin{figure}[!ht]
\begin{center}
\includegraphics[height=6cm,width=6cm]{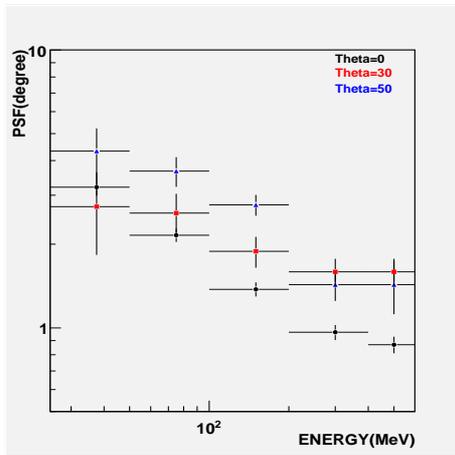}
\end{center}
\caption{Gaussian PSF versus energy for data.}
\label{PSFData}
\end{figure}
%%%%%%%%%%%%%%%%%%%%%%%%%%%%%%%%%%%%%%%%%%%%%%%%%%%%%%%%%%%%%%%%%%%%%%%%%%%%%%%%%%%%%%%%%%%%%%%%%%%

\section{Comparison of PSF for MC and real data {\it in phase}}

The PSF was evaluated in two different ways:
\begin{itemize}
\item a Gaussian fit plus a polynomial background, identifying the PSF with the Gaussian $\sigma$
\item the PSF is identified with the angular spread including 68\%\ of the events
\end{itemize}
The result of the 68\%\ estimation for the PSF for various $\theta$ versus $E_\gamma$ is shown 
in Fig.\ref{PSF68MC} for MC and in Fig.\ref{PSF68Data} for real data.\\
The result of the Gaussian estimation for the PSF for the same configurations is shown 
in Fig.\ref{PSFMC} for MC and in Fig.\ref{PSFData} for real data.\\
The 68\%\ PSF is significantly larger than the Gaussian PSF as expected in presence of background.\\
The data and MC Gaussian PSF are compatible with each others within the statistical errors.\\
For the 68\%\ PSF, the data show somehow larger values especially at low $E_\gamma$. That is likely
due to the low energy $\gamma$-ray background that is not adequately simulated. On the other hand, 
the Gaussian PSF should reflect more directly the quality of the GRID simulation, rather than the 
beam simulation.\\
An interpretation of these results is that the compatibility of the Gaussian PSF for data and MC 
represents a validation of the GRID simulation within the experimental requirements.

\section{Conclusions }

This paper presents some preliminary results of the calibration of the AGILE ST at the BTF 
of the LNF in 2005.\\
The setup is described in detail as well as the calibration requirements. We discussed the 
problems encountered in exploiting the PTS originally designed and a novel approach devised 
to circumvent those problems: the phase analysis.\\
We concentrated on the measurements of the PSF presenting two possible definitions: the Gaussian 
and the 68\%\ PSF.\\
The calibration results are compared with the MC simulations for a broad set of variables, showing 
good consistency with some poorer agreement for 68\%\ PSF mainly at low energies.\\
These results give confidence on the use of the MC simulation in the untested part of the 
$\gamma$-ray parameters (e.g. higher $E_\gamma$) especially in flight conditions, i.e. 
without low energy background, and in the measurement of detector parameters, like absolute 
efficiency and energy resolution, that are difficult to measure without exploiting the PTS 
information.

\section*{Acknowledgement}

We want to remember the memory of our coworker Dr. Fulvio Mauri who greatly contributed to all 
aspects of the calibration of AGILE and left us prematurely. 

\bibliographystyle{unsrt}
\bibliography{ricap}

\begin{thebibliography}{10}

\bibitem{agimis}
M.~Tavani et~al.
\newblock {\em Nucl. Instr. and Meth. A}, 588:52, 2008.

\bibitem{st1}
M.~Prest et~al.
\newblock {\em Nucl. Instr. and Meth. A}, 501:280, 2003.

\bibitem{minical}
C.~Labanti et~al.
\newblock {\em Nucl. Instr. and Meth. A}, 598:470--479, 2009.

\bibitem{ac}
F.~Perotti et~al.
\newblock {\em Nucl. Instr. and Meth. A}, 556:228, 2006.

\bibitem{superagile}
M.~Feroci et~al.
\newblock {\em Nucl. Instr. and Meth. A}, 581:728, 2007.

\bibitem{sttb}
G.~Barbiellini et~al.
\newblock {\em Nucl. Instr. and Meth. A}, 490:146, 2002.

\bibitem{geant}
R.~Brun et~al.
\newblock Geant detector description and simulation tool, 1993.

\bibitem{frukal}
R.~Fr\"uwirth.
\newblock {\em Nucl. Instr. and Meth. A}, 262:444, 1987.

\bibitem{giukal}
A.~Giuliani et~al.
\newblock {\em Nucl. Instr. and Meth. A}, 568:692--699, 2006.

\bibitem{gianotti}
F.~Gianotti et~al.
\newblock In Martin J.~L. Turner and Kathryn~A. Flanagan, editors, {\em Space
  Telescopes and Instrumentation 2008: Ultraviolet to Gamma Ray}, volume 7011,
  page 70113D. SPIE, 2008.

\bibitem{bulgarelli}
A.~Bulgarelli et~al.
\newblock In Martin J.~L. Turner and Kathryn~A. Flanagan, editors, {\em Space
  Telescopes and Instrumentation 2008: Ultraviolet to Gamma Ray}, volume 7011,
  page 70113C. SPIE, 2008.

\end{thebibliography}

\end{document}